\documentclass[nalefd]{achemso}
\setkeys{acs}{keywords = true}
\usepackage{amssymb}
\usepackage{graphicx}
\usepackage{epsfig}
\usepackage{multirow}
\usepackage{epstopdf}
\usepackage{gensymb}
\usepackage{nicefrac}
\usepackage{mathtools}
\usepackage{color}

\usepackage[left]{lineno}

\title{Large spatially-resolved rectification in a donor-acceptor molecular heterojunction}
\author{J. A. Smerdon}
\affiliation{Jeremiah Horrocks Institute of Mathematics, Physics and Astronomy, University of Central Lancashire, Preston, PR1 2HE, UK}
\author{N. C. Giebink}
\affiliation{Department of Electrical Engineering, The Pennsylvania State University, University Park, PA 16802, USA}
\author{N. P. Guisinger}
\affiliation{Center for Nanoscale Materials, Argonne National Laboratory, Argonne IL 60439, USA}
\author{P. Darancet}
\email{pdarancet@anl.gov}
\affiliation{Center for Nanoscale Materials, Argonne National Laboratory, Argonne IL 60439, USA}
\author{J. R. Guest}
\email{jrguest@anl.gov}
\affiliation{Center for Nanoscale Materials, Argonne National Laboratory, Argonne IL 60439, USA}

\date{today}


\begin{document}

\maketitle

\abstract{
We demonstrate that rectification ratios (RR) of $\gtrsim$250 ($\gtrsim$1000) at biases of 0.5~V (1.2~V) are achievable at the two-molecule limit for donor-acceptor bilayers of pentacene on C$_{60}$ on Cu using scanning tunneling spectroscopy and microscopy. Using first-principles calculations, we show that the system behaves as a molecular Schottky diode, with a tunneling transport mechanism from semiconducting pentacene to Cu-hybridized metallic C$_{60}$. Low-bias RR's vary by two orders-of-magnitude at the edge of these molecular heterojunctions due to increased Stark shifts and confinement effects.}
\\
\\
keywords: pentacene, fullerene, rectification, Schottky, STM, STS, DFT

\newpage

Since the first proposal for a molecular rectifier 40 years ago~\cite{Aviram:1974eh}, investigations into the transport properties of single-molecule devices have yielded demonstrations of molecular rectifiers, switches and other components~\cite{Tu:2008,Guisinger:2004hn,Zhao:2005ib,Jackel:2004ug,katsonis06-am,Kim14-pnas,Diao07-jap,metzger2015}.   However, despite theoretical models predicting that rectification ratios (RR) larger than 1000~\cite{taylor2002theory,Andrews08-jacs} are achievable for molecular diodes, experiments have reported much more modest RR's~\cite{metzger2015}. Furthermore, it has been difficult to unravel how nanoscale structure and local environment impact electrical transport through the molecular junctions that underlie operation of macroscale devices such as organic photovoltaic cells~\cite{Nelson:2009}. Following theoretical models \cite{taylor2002theory,Andrews08-jacs},  efforts towards the synthesis and characterization of more efficient molecular diodes can be divided into attempts to (1) increase the electron rich/poor characters of the donor/acceptor moieties~\cite{dhirani97-jcp},  (2) decrease their conjugation \cite{ashwell06-pccp,randel2014unconventional}, and  (3) imbalance their coupling to the electrodes~\cite{diezperez09-nc,kushmerick02-prl,selzer02-jpcb,guedon2012observation,batra2013tuning,Kim14-pnas}.  The experimental poor performance of these single-molecule diodes -- with the notable exception of environment-induced diodes~\cite{capozzi2015single} -- suggests that these physical parameters tend to be mutually exclusive in most molecular systems \cite{mujica02-cp,stokbro2003aviram}.   

In this Letter, inspired by thin-film organic photovoltaic devices~\cite{Tang:1986}, we simultaneously optimize  parameters (1-3) by assembling a bilayer heterojunction (HJ) of pentacene (Pn) and  C$_{60}$ - archetypal donor and acceptor molecules (respectively) - that weakly interact through van der Waals interactions \cite{Dougherty:2008hj,Dougherty09-apl}.  Using scanning tunneling microscopy (STM) and spectroscopy (STS), we resolve the structure of these HJ's, map transport properties and demonstrate RR's in excess of 1000.  We combine statistical current-voltage sampling methods used in single-molecule transport experiments with local STS to show a strong correlation between sub-nm-resolved rectification and statistically averaged rectification. Near the edge of the HJ, however, we show strong spatial dependence of the conductance channels (through d$I$/d$V$) and transport properties, demonstrating the importance of unraveling structure-function correlations in these systems. Using density functional theory (DFT), we show that this arrangement is a molecular analog to a Schottky diode.  Asymmetric coupling to electrodes is achieved through the strong coupling of C$_{60}$ with the Cu(111) substrate and the weak coupling of Pn with the STM tip.  Our heterojunctions (HJs) were formed by depositing a submonolayer coverage of Pn molecules on an annealed monolayer of C$_{60}$ on Cu(111). On Cu(111), C$_{60}$ adopts a $p$(4$\times$4) structure and hybridization of the molecular orbitals leads to a metallic electronic structure, associated with the strong broadening of the lowest unoccupied molecular orbital (LUMO) of C$_{60}$~\cite{wang2004rotation,PRB49:10717,PRB57:11939}. Annealing allows the C$_{60}$ to settle into the surface, displacing Cu atoms beneath each molecule and increasing the hybridization and metallicity to varying degrees \cite{Pai:2004bc,Pai10-prl}.  

Measurements were carried out in a ultrahigh-vacuum (UHV) variable-temperature (VT) STM operating with the sample maintained at 55~K.  The Cu(111) crystal was cleaned by sputtering with Ar$^+$ ions at 1~keV and simultaneously annealing at 900~K, with a final sputtering cycle at room temperature followed by a brief anneal at 900~K.  Pentacene and C$_{60}$ were deposited in the same UHV system using an organic molecular-beam evaporator at 510~K and 710~K, respectively.  The C$_{60}$ film was annealed to ~$\approx$570~K before Pn deposition.  Tungsten tips were prepared by electrochemical etching; when they occasionally adsorbed molecules during experiments, they were cleaned by $e$-beam heating to restore imaging and spectroscopy quality.  Bias voltage $V_B$ was applied to the sample.    Spectroscopy d$I$/d$V$ measurements were performed with a standard lock-in technique. The $I(V)$ and d$I$/d$V$  curves were obtained with a `grid' spectroscopy approach, where the measurements were obtained sequentially in a grid pattern over a 10~nm$\times$6.3~nm region of the sample.   The $I(V)$ current due to the shunt resistance ($R_S \approx 700$~G$\Omega$) of the instrument --measured with the tip far from the surface-- was subtracted from the measured $I(V)$ curves to obtain the $I(V)$ behavior of the molecular layers alone; this was critical for resolving the noise-limited current at our applied voltages.

\begin{figure*}[t]
\begin{center}

\includegraphics[width=8.46cm]{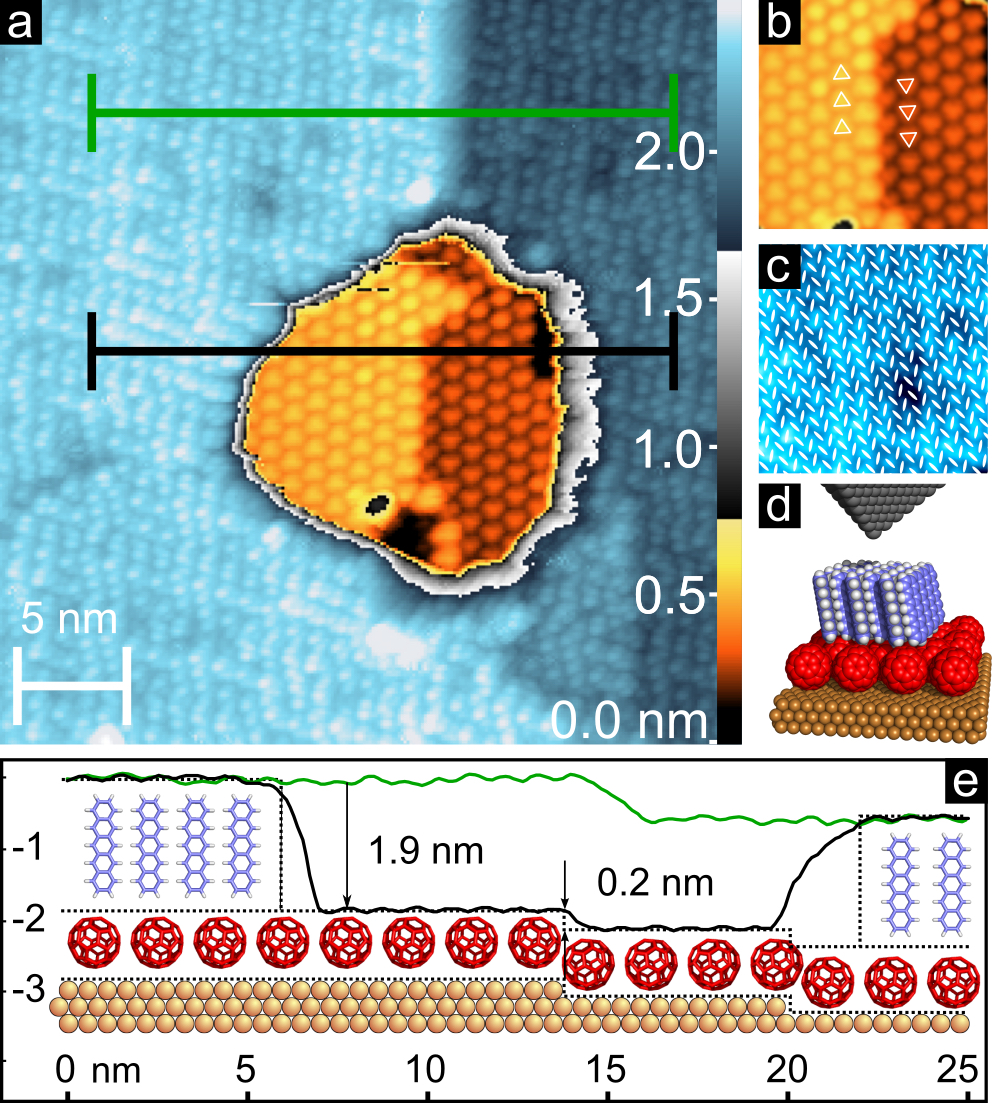}

\caption{Pentacene/C$_{60}$/Cu(111) system. (a) Scanning tunneling micrograph ($V_{B}=-0.5$ V, $I_T=80$ pA) showing Pn (blue) on C$_{60}$ (orange) on Cu(111) (substrate not visible). The color scale has been optimized to show fine structure in both layers, despite the large $z$-difference between them. (b) A 10$\times$10 nm area of the C$_{60}$ showing inverted orientations either side of the underlying substrate step. (c) A 10$\times$10 nm area of the Pn island calibrated according to the C$_{60}$ structure overlaid with a representation of the Pn herringbone structure. (d) A model of the system for illustration, showing an idealized substrate and tip. (e) Line profiles and a schematic of the area imaged in (a).}
\label{FigTopo}
\end{center}
\end{figure*}

In Fig.~\ref{FigTopo}(a), we show an STM image of our HJ.  Close-packed C$_{60}$ molecules remained uncovered in the center of the image as indicated by the threefold symmetry of the molecular orbitals (Fig.~\ref{FigTopo}(b)), providing the means to investigate the monolayer and the HJ simultaneously.  The line profiles in Fig.~\ref{FigTopo}(e) reveal that the measured step height from the C$_{60}$/Cu(111) to the Pn/C$_{60}$/Cu(111) HJ is $\approx 1.9$~nm, indicating that the Pn self-assembles into a  $\pi$-stacked structure standing vertically atop the C$_{60}$ monolayer; the herringbone arrangement of the Pn layer is overlaid in the image detail in Fig.~\ref{FigTopo}(c). The large-scale approximate periodicity visible in the HJ is a convolution of the underlying hexagonal topography of the C$_{60}$ monolayer and the distorted square unit cell of the Pn thin film structure \cite{Parisse:2006jz}.  A simple molecular model of the self-assembled HJ is shown in Fig.~\ref{FigTopo}(d).

\begin{figure*}[t]
\begin{center}
\includegraphics[width=17.78cm, angle=0]{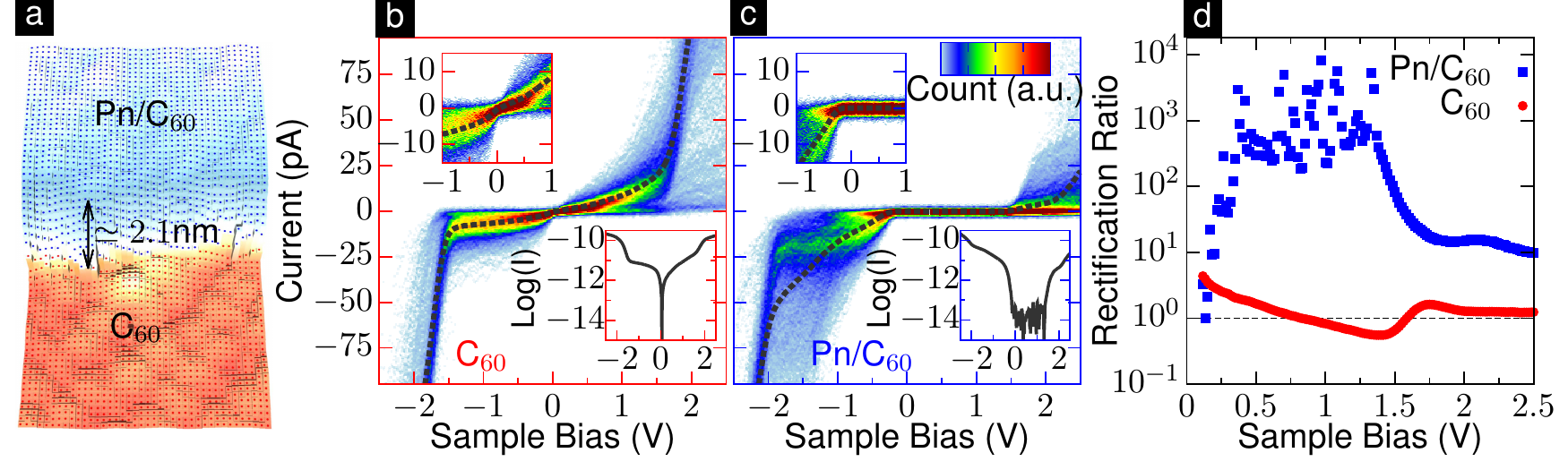}
\caption{Current-voltage ($I(V)$) characteristics of the  C$_{60}$/Cu(111) and Pn/C$_{60}$/Cu(111) regions. (a) Topograph of the scanned region ($10\times6.3$~nm$^2$), showing the C$_{60}$/Cu(111) (bottom) and  Pn/C$_{60}$/Cu(111) (top) regions differentiated by a $\simeq2.1$~nm step. $I(V)$ spectra recorded at every pixel are separated by height as indicated by red and blue dots and added to the histograms shown in (b) and (c), respectively.  The statistical distribution of $I(V)$ curves are thus shown as histograms for (b) C$_{60}$/Cu(111) and (c) Pn/C$_{60}$/Cu(111) regions; average $I(V)$ curves are shown by the superimposed dotted lines. Low-bias ($\left[ -1; 1\right]$V) linear plot and full-range bias ($\left[ -2.5; 2.5\right]$V)  logarithmic plot of the current are given in the top and bottom insets, respectively.   (d) Rectification Ratio ($\left| \nicefrac{I(-V)}{I(V)} \right| $) from each region is calculated from the average $I(V)$ curves and plotted. Positive (negative) biases correspond to electrons flowing from tip to sample (sample to tip). }
\label{FigHisto}
\end{center}
\end{figure*}

We explored the spatially-dependent $I(V)$ properties of this HJ by performing STS on a $10\times6.3$ nm$^2$ grid across the edge of the HJ, recording $I(V)$ curves at ($64 \times 40$) 2560 different points as indicated in Fig.~\ref{FigHisto}(a).  By `simultaneously' performing the measurements on both the HJ and the C$_{60}$  monolayer, we were able to verify that the same tip condition was maintained for both systems. Tip height at each point was established at the tunneling conditions $V_B=-2.5$~V and $I_{t}=200$~pA; no data was filtered from this set.  In addition to spatially-dependent $I(V)$ data, statistics were accumulated for these structures; $I(V)$ histograms were separated by their measured height (as indicated in Fig.~\ref{FigHisto}(a)) and plotted in Fig.~\ref{FigHisto}(b) and (c), corresponding to the C$_{60}$ monolayer and the HJ, respectively.  Average $I(V)$ curves are shown with the dotted lines. (Additional histograms are shown in S5).

The monolayer of C$_{60}$ (Fig.~\ref{FigHisto}(b)) on Cu shows a quasi-ohmic behavior for $-1$~V~$\leq V_B \leq 1$~V, with nonlinear increases in current arising at $\simeq+1$~V and $\simeq-1.5$~V as new conductance channels appear. Correspondingly, the rectification ratio ($RR= \left| \nicefrac{I(-V)}{I(V)} \right| $), computed from the average $I(V)$ and shown in Fig.~\ref{FigHisto}(d), is approximately unity up to biases of 2.5 V.

In contrast, the $I(V)$ characteristics in the Pn/C$_{60}$/Cu(111) HJ region are strongly non-ohmic: at positive bias, the current remains consistent with zero and under our sensitivity limit for biases as high as $\simeq1.3$~V (shown on a log scale in Fig.~\ref{FigHisto}(c), lower inset), while a negative bias $<$$-0.25$~V results in significant current.  The corresponding RRs (Fig.~\ref{FigHisto}(d)) are orders of magnitude larger than in the C$_{60}$/Cu(111) region with values approaching or $\gtrsim 1000$ for $0.4$~V~$\leq V_B \leq 1.3$~V, significantly larger than rectification reported for single-molecule systems~\cite{batra2013tuning,diezperez09-nc,kushmerick02-prl,selzer02-jpcb,randel2014unconventional}.  Interestingly, the preferred path for electrons - from acceptor to donor - is \emph{opposite} to the path found in most single-molecule rectifiers~\cite{metzger2015}. As explained below, this is due to the unique electronic structure of our system. 

The computed RRs $< 1.3$~V shown in Fig.~\ref{FigHisto}(d) are limited by instrument noise on measurements of zero current in the reverse direction, and are therefore a noisy indication of the lower bound of the actual RR.  In order to quantify the RR more comprehensively, we use the standard deviation of averaged current measurements between $-0.1$~V~$\leq V_B \leq 1.1$~V ($\sigma \approx 23$~fA, see SI) as a proxy for the detection sensitivity to establish confidence intervals for the RR; we find that, with 84\% confidence, the RR is $\gtrsim 250$, $\gtrsim 600$, and $\gtrsim 1000$ for $V_B$ of $0.5$~V, $0.9$~V, and $1.2$~V, respectively (complete map with various confidence levels is shown in the SI, Fig.~S6). 

\begin{figure*}[t]
\begin{center}
\includegraphics[width=17.78cm]{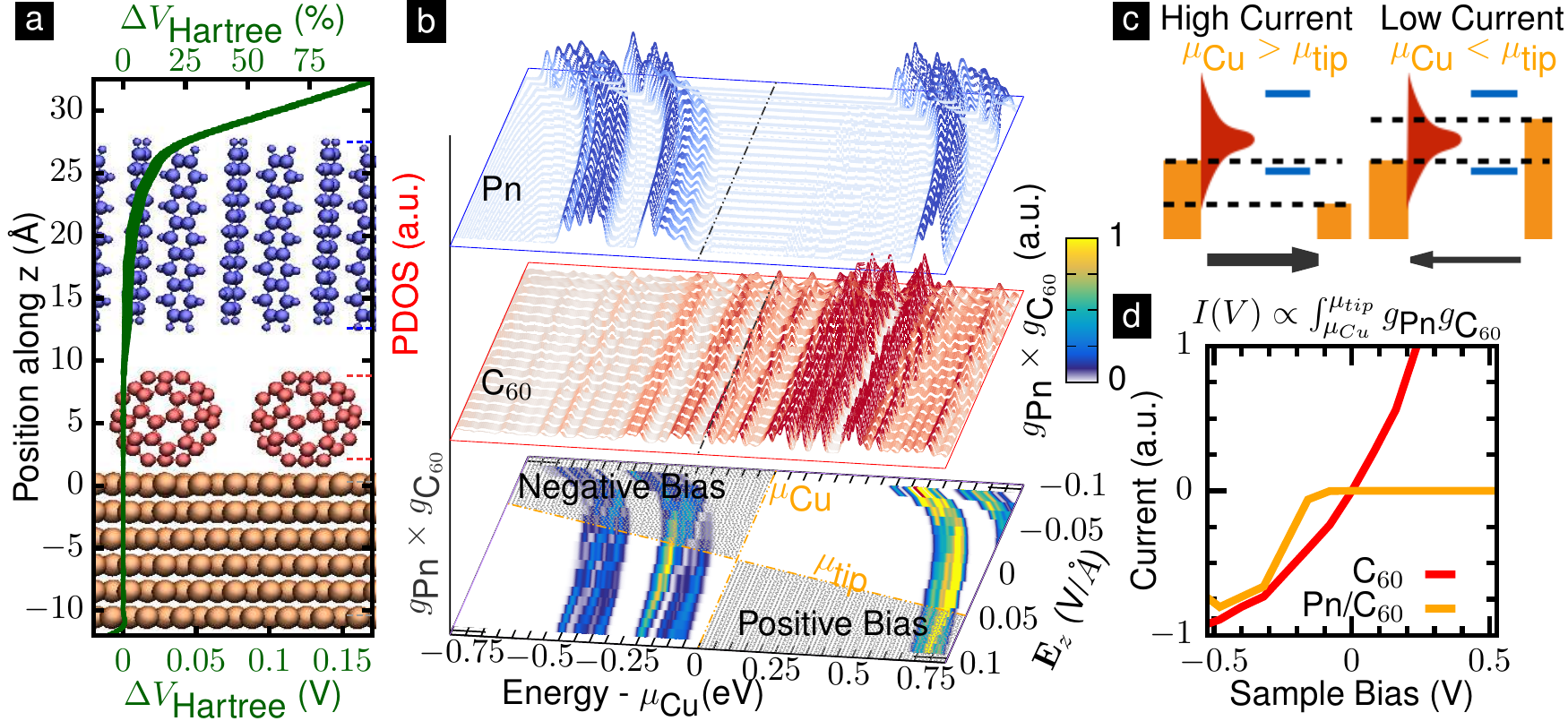}
\caption{Electronic structure and transport properties of Pn/C$_{60}$/Cu(111). (a) The potential drop across the slab for a tip-Pn distance of 5 \AA~as calculated in density functional theory (GGA-PBE) is superimposed on the structural model of the Pn/C$_{60}$/Cu(111) system. (b) Energy-resolved projected density of states (PDOS) in the Pn (top) and  C$_{60}$ (middle) layers. The product of the two PDOS is given in the bottom panel, with a shaded area corresponding to the bias window assuming a tip at $\simeq 0.5$~nm above the Pn layer. (c) Schematics of the level alignment in Pn/C$_{60}$/Cu(111). (d) Predicted tunneling current on top of Pn/C$_{60}$/Cu(111) (orange) and  C$_{60}$/Cu(111) (red) (using the product of  Pn and  C$_{60}$ PDOS for  Pn/C$_{60}$/Cu(111), and the PDOS of C$_{60}$/Cu(111) for the monolayer, see SI).  }
\label{FigTheory}
\end{center}
\end{figure*}

To understand the origins of this strong rectifying behavior, we perform first-principles density functional theory (DFT) calculations. Geometry optimization and electronic structure calculations are performed using density functional theory with the GGA functional of Perdew Burke Ernzerhof (PBE)~\cite{GGAPBE} as implemented in the SIESTA package~\cite{Siesta} (see the supplementary information (SI) for details). All structures are approximated by a slab comprising 6 layers of Cu with C$_{60}$ and Pn adsorbed on top. For all structures, atomic positions are optimized and forces on atoms are minimized to 0.04 eV/\AA. The bottom 4 layers of Cu are maintained in their bulk configurations during geometry optimization. In order to distinguish the electronic component of the polarizability from the bias-induced structural changes proposed in [\citenum{Andrews08-jacs,randel2014unconventional}], atomic positions are \emph{not} optimized when an external electric field is subsequently applied.  As the PBE functional does not account for van der Waals interactions, we considered two different Pn-C$_{60}$ distances; both produce qualitatively similar results. We use 3 Pn/C$_{60}$ in the unit cell (Fig.~\ref{FigTheory}(a)) as an approximant of the experimental structure of 3.2$\pm$0.2 Pn/C$_{60}$ measured in STM images (Fig.~\ref{FigTopo}(a), SI).  The bias is modeled by an external electric field applied in the out-of-plane direction. The relaxed atomic positions of these structures along with their changes in Hartree potential at finite electric fields are given in the SI. 

By considering the difference in Hartree potentials at zero and finite electric fields, we find that the bias drops almost exclusively in the vacuum and Pn regions. While strong screening is expected for a metal such as Cu, the strong polarizability of the C$_{60}$ layer is an emergent property of the strong coupling of C$_{60}$ to Cu(111) and its subsequent partial metallization \cite{Pai:2004bc,wang2004rotation}; in contrast, the semiconducting Pn layer can only partially screen the field.  In this way, the molecular HJ parallels a metal/semiconductor junction (\textit{i.e.}\ a Schottky diode).

This Schottky diode analogy is further validated by examining the density of states projected on the molecular orbitals (PDOS) of each layer (Fig.~\ref{FigTheory}(b)).  The chemical potentials of the tip and Cu are related to the experimental voltage bias (applied to the sample) as $e V_B=\mu_{tip}-\mu_{Cu}$.  The Pn, decoupled from the substrate, shows a clear gap in its PDOS (blue, top) at $E_F$, with a $\simeq0.1$-eV-broad peak corresponding to the highest occupied molecular orbital (HOMO) of Pn located 0.15 eV below $E_F$. In contrast, the C$_{60}$ layer has significant PDOS (red, middle) at $E_F$, which arises from the strongly broadened LUMO at 0.4 eV above $E_F$. 

As we will now show, the large rectification is a direct consequence of this unusual electronic structure. Specifically, a minimal tunneling model assuming weak coupling between the molecules and the STM tip in combination with the DFT electronic structure is sufficient to explain our findings.  Under this mechanism, the electronic transmission is proportional to the product of the PDOS on Pn and C$_{60}$, plotted in Fig \ref{FigTheory}(b). For $V_B>0$, electrons must tunnel through the gap of Pn, leading to a transmission probability near zero. For $V_B<0$, electrons can tunnel into the finite density of states of C$_{60}$ hybridized LUMO and then into the Pn HOMO inside the bias window, resulting in a sharp increase in current as the Pn HOMO enters the bias window. In contrast, in the C$_{60}$/Cu(111) region, electrons must tunnel through the strongly-broadened LUMO of C$_{60}$, resulting in no significant rectification. This simple model, which neglects the hybridization between the tip and the HJ and is based on the Kohn-Sham level-alignment, is not expected to provide a quantitative picture; it nevertheless leads to a remarkable qualitative agreement in the $I(V)$ as shown in Figure \ref{FigTheory}(d).  It is moreover consistent with the weak rectification of C$_{60}$/Pn/Cu(111)~\cite{Yang:2007by} as the strong C$_{60}$-Cu(111) coupling is necessary to obtain large rectification.

\begin{figure*}[ht!]
\begin{center}
\includegraphics[width=17.78cm]{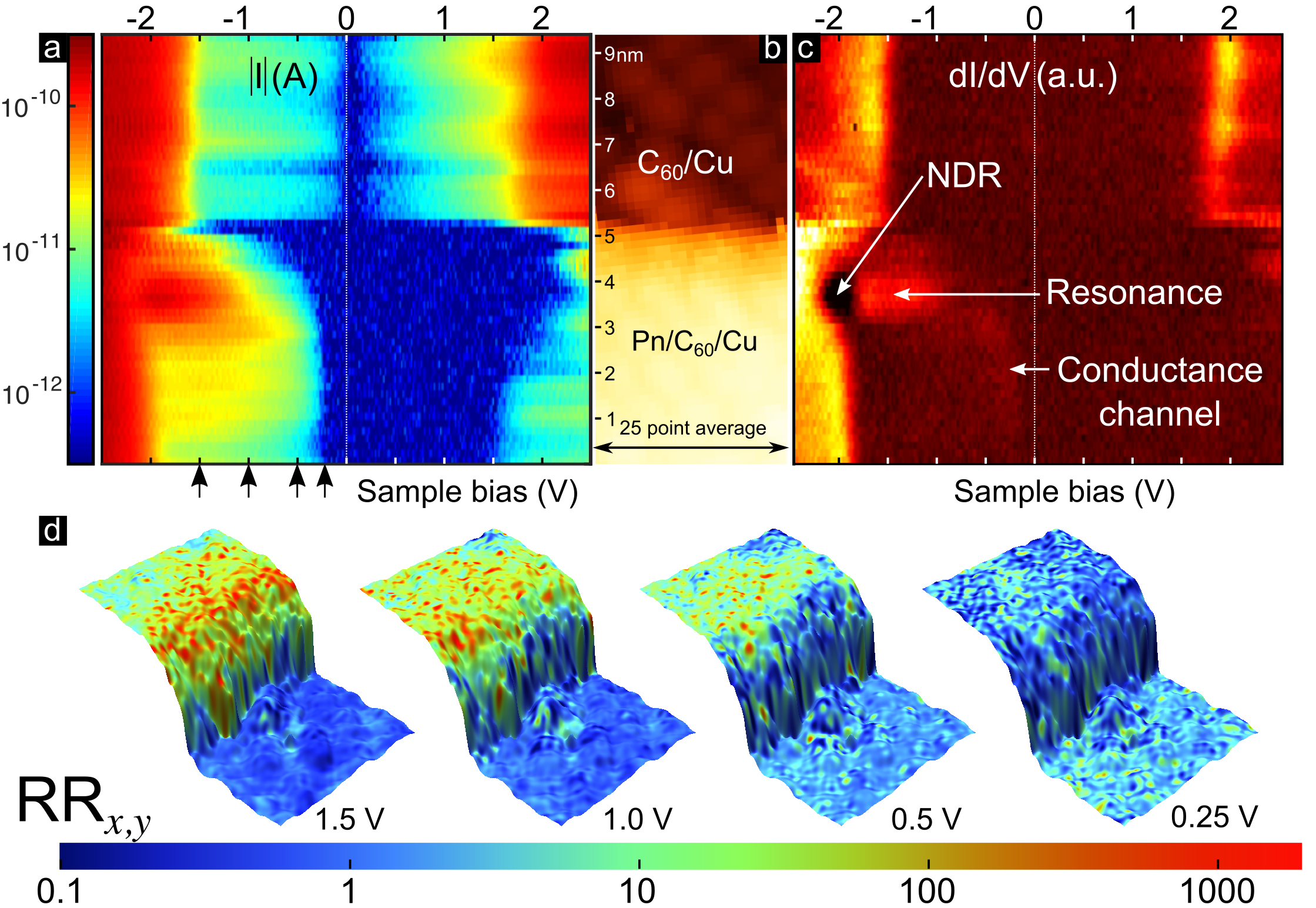}
\caption{Spatially-resolved (a) $I(V)$ and (c) d$I$/d$V$ spectrograms, where the spatial dimension parallel to the HJ edge has been averaged  over 25 locations across a line parallel to the island edge as indicated in the corresponding topograph shown in (b).  The abrupt change of electronic structure is clearly visible at the HJ edge. Some features in the differential conductance are annotated in (c). (d) 10 nm $\times$ 6.3 nm topographs color-coded with the local RR at the specified voltage (and indicated by arrows in (a)).}  
\label{Figdidv}
\end{center}
\end{figure*}

This mechanism is further elucidated by the spatially-resolved maps of $I(V)$ and d$I$/d$V$ obtained in STS and presented in Fig.~\ref{Figdidv}(a) and Fig.~\ref{Figdidv}(c), respectively, where curves taken on a line parallel to the HJ edge are averaged as indicated in  Fig.~\ref{Figdidv}(b).  Independently of the distance to the edge, the C$_{60}$/Cu region shows ohmic behavior and the Pn/C$_{60}$/Cu HJ has a large region of zero current (dark blue region, at the level of instrument noise) for $-0.25$~V~$\leq V_B \leq 1.5$~V. Rectification in Pn/C$_{60}$/Cu HJ arises as a resonance seen in the  d$I$/d$V$ map enters the bias window at $-0.25$~V (in agreement with the DFT prediction of Pn HOMO at $-0.15$~eV).

Moreover, the energy of this Pn HOMO resonance clearly varies as a function of distance to the edge and considerably impacts the transport properties of the HJ within a few nanometers of the edge (Fig.~\ref{Figdidv}(c)).  Near the HJ edge, the HOMO shifts to more negative $V_B$ (or higher negative energy); we attribute this behavior to a combination of confinement energy of the HOMO and tip-induced Stark shift due to the reduced screening at the HJ edge. The Stark shift towards lower energy at negative bias is also consistent with Figure \ref{FigTheory}(b).  As the Pn HOMO shifts towards the C$_{60}$ HOMO, there is a large peak in differential conductance with a concomitant peak in $I(V)$, followed  by a {\it drop} in current (negative differential resistance (NDR)) -- a result which is expected if the HOMOs of Pn and C$_{60}$ are tuned through resonance with one another~\cite{Guisinger:2004hn,Tu:2008,perrin2014large}.  Note that this spatially-varying response accounts for much of the divergence in the $I(V)$ histograms for the HJ observed in Fig.~\ref{FigHisto}(c).
 
The corresponding spatial maps of the RR are plotted in Fig.~\ref{Figdidv}(d) for the specified biases (see SI for comprehensive RR maps).  Over the C$_{60}$ monolayer, the local RR is  $<5$ over the entire bias range.  In contrast, we observe significant variation in the local rectification in the HJ within  $\simeq$2 nm of the edge of the HJ as the energy of the Pn HOMO state is shifting.  At low voltages ($\approx 0.5$~V), the RR increases from approximately unity at the HJ edge to $\sim 100 \times$ larger $\simeq$2 nm inside the HJ.  Conversely, at higher voltages ($\simeq 1.5$~V), the RR is largest right at the edge of the HJ and decreases inside the HJ.  This behavior can be traced to the redistribution of the spectral weight of the Pn HOMO near the edge of the HJ observed in Fig.~\ref{Figdidv}(c). These spatial maps of the RR imply that rectification will occur at higher biases in the limit of a single Pn molecule on  C$_{60}$/Cu(111). 

In conclusion, we have observed for the first time a non-covalent, self-assembled molecular HJ consisting of single layers of molecules with an observed RR on the order of 1000 at biases less than 500~mV. This observation is limited by the current noise floor of our apparatus rather than by the diode itself. This arrangement constitutes a molecular Schottky diode, with $I(V)$ characteristics consistent with a tunneling mechanism. Moreover, we have spatially resolved the rectification ratio and show that it varies by two orders of magnitude on a nm length scale near the edge of the HJ. The simplicity of the arrangement indicates that these findings may have implications for larger scale molecular electronics -- such as organic photovoltaics -- where this system could act as an efficient electron-blocking layer.

\section{Acknowledgments}

Use of the Center for Nanoscale Materials, an Office of Science user facility, was supported by the U. S. Department of Energy, Office of Science, Office of Basic Energy Sciences, under Contract No. DE-AC02-06CH11357.  Primary support for this work was provided by the Department of Energy Office of Basic Energy Sciences (SISGR Grant DE-FG02-09ER16109).  J.A.S. acknowledges support through the UK Science \& Innovation Network and Department for Business, Innovation, and Skills.  The authors acknowledge the technical assistance of B.~L. Fisher and discussions with M. Bode.

\section{Supporting Information}

The following material is available free of charge via the Internet at http://pubs.acs.org: 
Experimental details and more data on the C$_{60}$/Cu(111) and Pn/C$_{60}$/Cu(111) structures. 
Full $I(x,y)$ and RR vs $V$ maps for the presented dataset. 
Zoomed-in linear and log histogram representations. 
Noise analysis and RR confidence intervals.
Complete set of RR maps.
Additional RR maps of a dataset with STS feedback stabilized in the reverse direction of the diode. Computational details for DFT simulations. DFT results using different geometries. Two-site model. Optimized coordinates of the slabs. 
This material is available free of charge via the Internet at http://pubs.acs.org.

\section{Notes} 

The authors declare no competing financial interest.

\providecommand{\latin}[1]{#1}
\providecommand*\mcitethebibliography{\thebibliography}
\csname @ifundefined\endcsname{endmcitethebibliography}
  {\let\endmcitethebibliography\endthebibliography}{}

\end{document}